\documentclass[prd,aps,amssymb,amsmath,tightenlines,nofootinbib]{revtex4}
\usepackage{graphicx,epsfig}
\usepackage{slashed}
\usepackage[colorinlistoftodos]{todonotes}
\usepackage{appendix}
\usepackage{hyperref}

\newcommand{\ee}{\end{equation}}

\newcommand{\be}{\begin{equation}}

\def\cPT{$\mathcal{PT}$}

\usepackage{subcaption}

\usepackage{tikz}
\usetikzlibrary{calc,patterns,angles,quotes}

\usepackage{amsmath}

\begin{document}

\preprint[{\leftline{KCL-PH-TH/2022-{\bf 57}}

\title{\boldmath New classes of solutions for Euclidean scalar field theories}

\author{Carl M. Bender$^a$}
\author{Sarben Sarkar$^b$}

\affiliation{$^a$Department of Physics, Washington University, Saint~Louis,
Missouri 63130, USA}
\affiliation{$^b$Theoretical Particle Physics and Cosmology, King's College London, Strand, London, WC2R 2LS, UK}

\begin{abstract}
This paper presents new classes of exact radial solutions to the nonlinear ordinary differential equation that arises as a saddle-point condition for a Euclidean scalar field theory in $D$-dimensional
spacetime. These solutions are found by exploiting the dimensional consistency of the radial differential equation for a single {\it massless} scalar field, which allows one to transform to an autonomous equation. For massive theories the radial equation is not exactly solvable but the massless solutions provide useful approximations to the results for the massive case. The solutions presented here depend on the power of the interaction and on the spatial dimension, both
of which may be noninteger. Scalar equations arising in the study of conformal invariance fit into this framework and classes of new solutions are found. These solutions exhibit two distinct behaviours as $D\to2$ from above.
\end{abstract}

\maketitle

\section{Introduction}
\label{s1}
Tunneling is a process allowed in quantum physics but not in classical physics. Tunneling amplitudes in quantum mechanics can be calculated approximately by using WKB techniques \cite{R15, Bender:1969si}. In quantum field theory the transition between a metastable (false) vacuum state and the true ground state (vacuum) is a tunneling process. This process can be studied in field theory in spacetime dimension $D$ by using the path integral representation of the Euclidean partition function $Z$. Here, we consider a field theory for a scalar field $\phi$ for which $Z$ is given by the path integral
$${\cal Z}=N\int\cal{D}\phi\,\exp(-S[\phi]/\hbar).$$ 

In the semiclassical limit \cite{Coleman:1985rnk,PhysRevD.15.2929,R34cc} $\cal Z$ is approximated by calculating fluctuations around the classical solution $\phi_c$ of the equation of motion
$\frac{\delta S[\phi]}{\delta\phi(x)}=0$
for the scalar field $\phi$. The usual form of $S[\phi]$ is 
\be\label{e1}
S[\phi]=\int d^4x\big[ \tfrac{1}{2}(\partial_\mu \phi)^2+\tfrac{1}{2}m^2 \phi^2+\tfrac{g}{4} \phi^4\big]  
\ee
for which the equation of motion is 
$$\partial^2\phi-m^2\phi -g\phi^3=0.$$
For the massless case $m^2=0$ there is a well-known explicit solution to this equation known as the {\it bounce}~ \cite{Fubini1976-qg, Lipatov:1976ny, Coleman:1985rnk}, but for nonzero $m^2$ there are no known explicit solutions.

This paper presents new classes of solutions for the case in which the Euclidean spacetime dimension $D$ is noninteger and the power of the scalar interaction may differ from $4$. Generalising the power of the interaction is natural in the context of the $\delta$-expansion \cite{Bender:1989vv}  approach to quantum field theory and in the field of \cPT-symmetric quantum field theory \cite{R3b, Felski:2021evi, Croney:2023gwy, R3a, R4, PRD98.125003}. Thus, instead of the action \eqref{e1} we consider
the more general action
$${\cal S}[\phi]=\int d^Dx\,\big[\tfrac{1}{2}(\partial_\mu\phi)^2+\tfrac{1}{2}m^2\phi^2+\lambda\phi^{2+\epsilon}\big],$$
where $\epsilon$ is a real parameter. The equation of motion for this more general action is 
\be\label{EE3}
\partial^2\phi-m^2\phi-(2+\epsilon)\lambda \phi^{1+\epsilon}=0.
\ee
We cannot solve this nonlinear partial differential equation, but following the work of \cite{Coleman:1977th}, we shall assume that the classical instantons of least action are spherically symmetric.
Consequently, we will replace this differential equation by the spherically-symmetric ordinary differental equation
\be\label{EE4}
\phi''(r)+\tfrac{D-1}{r}\phi'(r)-m^2\phi(r)-(2+\epsilon)\lambda\phi^{1+\epsilon}(r)=0.
\ee

In this paper we apply techniques taken from the theory of ordinary differential equations to solve the massless version of the radial equation of motion \eqref{EE4} and discover infinite numbers of new solutions. We discuss the limiting behavior of two classes of these solutions as $D$ approaches $2$ from above. We clarify the role of scale invariance in restriciing the class of solutions with finite action. We also show how the notions of boundary-layer theory can be applied to justify the use of these massless solutions to approximate the contributions of massive solutions.

\section{The role of scale invariance}\label{s2}
Consider the Euclidean action ${\cal S}_m[\phi]=\int d^Dx\,{\cal L_m}[\phi(x)]$ derived from the Lagrangian
\begin{equation}
\label{ee1}
{\cal L_m}(\phi)=\tfrac{1}{2}\partial_\mu\phi\partial^\mu\phi+\tfrac{1}{2}m^2 \phi^2-\lambda m^{2+\epsilon-\epsilon D/2}\phi^{2+\epsilon},
\end{equation}
where $\lambda$ is a dimensionless coupling constant. (In 
\cPT-symmetric models
$\lambda$ may be complex \cite{Felski:2021evi}). 
Scale invariance requires that a solution $\phi_c(x)$ of the equations of motion satisfy
\begin{equation}
\label{ee2}
\frac{d}{d\eta}S[\phi_c /\eta]\big|_{\eta=1}=-\int d^Dx\,\left([\partial_\mu\phi_c(x)]^2+m^2\phi^2_c(x)\right)+(2+\epsilon)\lambda m^{2+\epsilon-\epsilon D/2}\int d^Dx\,\phi^{2+\epsilon}_c(x)=0.
\end{equation} 
Hence 
\begin{equation}
\label{ee2a}
(2+\epsilon )S[\phi_c]=\tfrac{1}{2}\epsilon\int d^Dx\ \big[\partial_\mu \phi_c(x)\big]^2+\tfrac{1}{2}{\epsilon\ m^{2+\epsilon-\epsilon D/2}}\int d^Dx\,[\phi_c(x)]^2\,. 
\end{equation}

Similarly $\frac{d}{d\eta} S[\phi_c(\eta x)]\big |_{\eta=1}=0$, which implies that
\begin{equation}
\label{ee4}
S[\phi_c]=\tfrac{1}{D}\int d^Dx\,\left[\partial_\mu\phi_c(x) \right]^2>0. 
\end{equation}
Thus, assuming that $S[\phi_c]$ is finite, we deduce from \eqref{ee2a} and \eqref{ee4} that
\begin{equation}
[(\tfrac{1}{2}D-1)\epsilon-2]S[\phi_c]=\tfrac{1}{2}\epsilon m^{2+\epsilon-\frac{1}{2}\epsilon D}\int d^Dx\,[\phi_c(x)]^2.
\end{equation}
We conclude that for $\epsilon=\tfrac{4}{D-2}$ there are only solutions with $m=0$. 
 
\section{The role of mass}
Explicit massive solutions cannot be found for $D\ne 1$. We redefine the field variable as $\phi_c(x)= \alpha\,\varphi(mx)$, where 
$\alpha=m^{D/2-1}
(2+\epsilon)^{-1/\epsilon}\lambda^{-1/\epsilon}$.
In terms of this new field the classical equation of motion is
$$-\varphi''(mx)+\varphi(mx)-\varphi^{1+\epsilon}(mx)=0.$$
As noted earlier, the smallest-action solution is radially symmetric in the variable $r=|mx-mx_0|$, where $x_0$ is an arbitrary point in Euclidean spacetime. The equation of motion now reads
$$-\varphi''(r)-\tfrac{D-1}{r}\varphi'(r)
+\varphi(r)-\varphi^{1+\epsilon}(r)=0.$$

This equation does not have solutions that obey the boundary conditions for a bounce~\cite{Coleman:1985rnk, Derrick1964-ff}, but a solution may have a bounce-like core. Indeed for $\epsilon>0$ and small $\phi$ and using {\it dominant balance}, we can solve the equation
\be\label{ee1}
-\varphi''(r)-\tfrac{D-1}{r}\varphi'(r)+ \varphi(r)=0
\ee
for large $r$, which has the independent solutions $\left\{r^{(2-D)/2}I_{(D-2)/2} (r),~r^{(2-D)/2}K_{(D-2)/2}(r)\right\}$ \cite{R15}. We are interested in finite actions and so we choose $K_{(D-2)/2}(r)$ since it falls off exponentially with $r$; the function $I_{(D-2)/2}(r)$ increases exponentially with $r$ and will not be considered. This large-$r$ {\it outer} solution can be matched~\cite{R15} with the small-$r$ {\it inner} solution, which we obtain in the massless approximation used in the next section. The scale of the core of this solution in $x$ is of order $1/m$. For small $r$ the $\frac{D-1}{r}\frac{d}{dr}$ term in \eqref{ee1} is large compared with terms of order $1$. Ignoring these smaller terms, we get the massless equation considered in the remainder of this paper.\footnote{It was shown 
in~\cite{Coleman:1977th} that massive solutions exist in our case for $D<D_c=(4+2\epsilon)/\epsilon$.}

The concepts used above originate in boundary-layer analysis~\cite{R15}. We treat the central core of a soliton as the {\it outer region} and the edges of the soliton as the boundary layer or {\it inner region}. In the inner region the soliton varies rapidly and decays exponentially to zero on a scale determined by the mass term. In the core region the soliton is slowly varying and this region makes the principal contribution to the action; in this core region the mass term in the Lagrangian is neglected. For massless classical solutions there are quantum fluctuations that radiatively generate masses \cite{Andreassen:2016cvx, Coleman:1985rnk, Coleman:1973jx}. Consequently massive solutions may not have a dominant role in the physics of tunneling.

\section{Solutions of the massless equation}
The massless field theory for a scalar field $\phi(x)$, where $x\in
\mathbb{R}^D$, is given by the Lagrangian
\begin{equation}
\label{e1}
{\mathcal L}(\phi)=\tfrac{1}{2}\partial_\mu\phi\partial^\mu\phi+\lambda\phi^{2
+\epsilon}.
\end{equation}
The semiclassical evaluation of the path integral  
\begin{equation}
\label{e2}
Z=\int D\phi\,\exp\left[-\int d^Dx\,{\mathcal L}(\phi)\right]  
\end{equation}
requires the solution of the saddle-point equation
\begin{equation}
\label{e3}
\partial_\mu\partial^\mu\phi-(2+\epsilon)\lambda\phi^{1+\epsilon}=0.
\end{equation}
(The cases $\epsilon=-1,\, -2$ are trivial and will not be considered.} 

This equation is invariant under a translation of $x$ by $x_0$; in terms of $r=|x-x_0|$, the radially-symmetric solution of \eqref{e3} is found by solving
\begin{equation}
\label{e4}
\phi''(r)+\tfrac{d}{r} \phi'(r)-\lambda(2+\epsilon)\phi^{1+\epsilon}=0
\end{equation}
where $d=D-1$. Ordinarily, the boundary conditions used in studies of vacuum decay~\cite{Coleman:1985rnk} are $\tfrac{d\phi}{dr}\big\vert_{r=0}\!=0$ and 
$\lim_{r\to\infty}\phi(r)\to\phi_0$.

Note that \eqref{e4} is {\it dimensionally consistent}; that is, it is
invariant under a scale change in $r$ \cite{R15}. For dimensional consistency we make a change of variable from $\phi(r)$ to $f(r)$: $\phi(r)=f(r)r^{-2/\epsilon}$. This leads to an {\it equidimensional} equation (an equation that is invariant under $r\to\alpha r$) \cite{R15}:
\begin{equation}
\label{e6}
r^2f''(r)+(d-\tfrac{4}{\epsilon})rf'(r)+\tfrac{2} {\epsilon^2}[2-(d-1)
\epsilon]f(r)-\lambda(2+\epsilon)f^{1+\epsilon}(r)=0.
\end{equation}

We make the further change of variable $r=e^t$, which is equivalent to substituting
$r\frac{d}{dr}=\frac{d}{dt}$. This gives a second-order {\it autonomous} equation in which the independent variable does not appear explicitly:
\begin{equation}
\label{e7}
\epsilon^2(\tfrac{d^2}{dt^2}-\tfrac{d}{dt})u(t)+\epsilon(d\epsilon-4)
\tfrac{d}{dt}u(t)+2[2-(d-1)\epsilon]u(t)-\lambda\epsilon^2(2+\epsilon)u^{1+
\epsilon}(t)=0,
\end{equation} 
where $u(t)\equiv f(r)$. 
Equation \eqref{e7} is an autonomous equation in $t$. This suggests substituting a new function $F(u)$ such that
\begin{equation}
\label{e8}
u'(t)=F(u),
\end{equation}
which reduces the order of the differential equation by one~\cite{R15}.

In terms of $F(u)$ we rewrite $\eqref{e7}$ as
\begin{equation}
\label{e9}
F(u)F'(u)+(d-\tfrac{4}{\epsilon}-1)F(u)+\tfrac{2}
{\epsilon^2}[2-(d-1)\epsilon]u-\lambda(2+\epsilon)u^{1+\epsilon}=0,
\end{equation}
which is a {\it first-order} equation.

\section{Classes of instanton solutions}

\subsection{Solutions with nonpolynomial $F(u)$}
\label{nonpoly}
If $d=\tfrac{4}{\epsilon}+1$, \eqref{e9} is easy to solve:
\begin{equation}
\label{e10}
F(u)=\pm\sqrt{2}(2\epsilon^{-2}u^2+\lambda u^{\epsilon+2}+c)^{1/2}, 
\end{equation}
where $c$ is a constant. This leads to 
\begin{equation}
\label{e11}
(2\epsilon^{-2}u^2+\lambda u^{\epsilon +2}+c)^{-1/2}du
=\pm {2}^{1/2}dt.
\end{equation}
It is easiest to do the $u$ integration in \eqref{e10} for $c=0$; if $c\neq0$
one can do the integration in terms of elliptic functions. For the choice $c=0$,
we find a class of solutions
\begin{equation}
\label{e12}
\phi(r)=(\tfrac{2\rho^2}{\lambda\epsilon^2})^{1/\epsilon}(\tfrac{1
}{r^2+\rho^2})^{2/\epsilon}, 
\end{equation} 
where $\rho$ is a scale which arises from a constant of integration.

In a \cPT-symmetric theory $\lambda=gi^\epsilon$, where $g>0$, so $\phi$ is complex in general. For $\epsilon=4$, $d=2$ ($D=3$). For $\epsilon=2$, $d=3$ ($D=4$) the
solution is real for positive $\lambda$ and it is the well-known bounce solution studied in false vacuum decay \cite{Coleman:1985rnk}. The bounce solution for $\epsilon=4$, $d=2$ is not typically discussed and to the best of our knowledge is new. The general solution for $\epsilon=\frac{4}{d-1}$ is new. Here, $d$ may be
noninteger.

\subsection{Solutions with polynomial $F(u)$}
If the linear term in $F$ in \eqref{e9} is present, the method above does not
apply. However, a different method of solution can be devised for a special
family of $d$. This is an entirely new family of solutions that can be considered as inner solutions for solitons. 

The case $\epsilon=2$ in \eqref{e9} leads to
\begin{equation}
\label{e13}
F(u)F'(u)+(d-3)F(u)+(2-d)u-4\lambda u^3=0.
\end{equation}
We substitute $F(u)=a_0+ 
a_1u+a_2u^2$ into
\eqref{e13}.\footnote{No new solutions are obtained by generalising to Laurent polynomials. If $F$ is a polynomial of degree $p$, the term $F(u)F'(u)$, which is of order $2p-1$, needs to be $\epsilon+1$ for $F$ to satisfy~\eqref{e9}.}
Matching powers of $u$, we obtain four simultaneous equations:
\begin{align}
a_0a_1+(d-3)a_0&=0,\nonumber\\
a^2_1+2a_0a_2+(d-3)a_1+2-d&= 0,\nonumber\\
3a_1a_{{}_{2}}+(d-3)a_2&=0,\nonumber\\
a^2_2-2\lambda &=0.\label{e14d}
\end{align}

The key feature of these equations is that the number of equations exceeds the number of variables. (Note that $d=3$ is a special case discussed in Sec.~B1,
so we consider $d\neq3$.) On solving \eqref{e14d} in sequence we obtain two solutions:
$$d=0~{\rm with}~a_0=0,~ a_1=1,~a_2=\pm\sqrt{2\lambda};$$
$$d=3/2~{\rm with}~a_0=0,~ a_1=1/2,~a_2=\pm\sqrt{2\lambda}.$$
  
The case $\epsilon=4$ in \eqref{e9} leads to
\begin{equation}
\label{e15}
F(u)F'(u)+(d-2)F(u)-6\lambda u^5+\tfrac{1}{4}(3-2d)u=0.
\end{equation}
We substitute $F(u)=a_0 +a_1u+ a_2u^2+a_3u^3$ into \eqref{e15}. Then, matching powers of $u$, we obtain six equations in four unknowns:
\begin{align}
(d-2+a_1)a_0 &=0,\nonumber\\
\tfrac{1}{4}(3-2d)-(2-d)a_1+a^2_1+2a_0a_2&=0,\nonumber\\
(d+3a_1-2)a_2+3a_0a_3&=0,
\nonumber\\
\left(d+4a_1-2\right)a_3 +2a^2_2&=0,\nonumber\\
5a_2a_3&=0,\nonumber\\
a^2_3-2\lambda &=0. \label{e16f}
\end{align}

The solution to these equations is
$$a_0=0,\quad a_1=\tfrac{1}{2}- \tfrac{d}{4},\quad a_2=0,\quad a_3=\pm\sqrt{2\lambda}.$$
From the second of these equations we deduce that either $d=0$ or $d=\tfrac{4}{3}$; that is, $a_1=\tfrac{1}{2}$ or $a_1=\tfrac{1}{6}$.
(We consider the cases $\epsilon=6,\,8$ in the Appendix.) In these examples there are always $\epsilon+2$ equations and $\tfrac{ \epsilon}{2}+2$ unknowns, so there are always {\it more equations than unknowns}. It is therefore quite remarkable that solutions exist.

The following pattern has emerged and has been explcitly verified: For $\epsilon=2n~(n=1,2,3,\ldots)$ the solutions to \eqref{e9} require that either $d=0$ or $d=\tfrac {n+2}{n+1}$. For $d= \frac{n+2}{n+1}=\tfrac {\epsilon+4}{\epsilon+2}$ we have 
\begin{equation}
\label{e24}
F(u)=a_1u+a_{n+1}u^{n+1},
\end{equation}
where $a_1=\frac{1}{n(n+1)}$ and $a_{n+1}=\pm\sqrt{2\lambda}$. For $d=0$ we have $a_1=\frac{1}{n}$  and $a_{n+1}=\pm\sqrt{2\lambda}$. From \eqref{e8} for $\epsilon=2n$ we have
\begin{equation}
\label{e25}
\frac{du}{dt}=\begin{cases}
\frac{1}{n}u\pm\sqrt{2\lambda} u^{n+1}~&~ (d=0),\\
\frac{1}{n(n+1)}u\pm\sqrt{2\lambda}u^{n+1}~&~ \left[d=(n+2)/(n+1)\right].
\end{cases}
\end{equation}

To summarise, for $d=0$ a family of solutions is 
\begin{equation}
\label{e26}
u=\left(\frac{\alpha\exp(t)}{1+\sqrt{2\lambda}\,\alpha n\exp(t)}\right)^{1/n},
\end{equation}
where the constant $\alpha>0$, and the corresponding $\phi$ is given by
\begin{equation}
\label{e27}
\phi=\left( {}\frac{\alpha }{1+\sqrt{2\lambda }\alpha n\exp(t)}\right)^{1/n}.  
\end{equation}
 
A corresponding set of solutions arises if the negative sign is taken in \eqref{e25}. For $d=\tfrac{n+2}{n+1}$ a family of solutions is
\begin{equation}
\label{e28}
\phi=\left(\frac{\alpha\exp\left(-\frac{nt}{n+1}\right)}{1+\sqrt{2\lambda}\,n (n+1)\alpha\exp\left(
\frac{t}{n+1}\right)}\right)^{1/n}.  
\end{equation}

Note that $d=0$ corresponds to $D=1$ and \eqref{e1} resembles quantum
mechanics with a power potential. If $n\to\infty$ this potential becomes a
square well. The above solutions may have implications for strong-coupling behaviour in quantum mechanics in unstable or $\cal {PT}$-symmetric potentials.

Since $d=\frac{n+2}{n+1}$ corresponds to $D=2+\frac{1}{n+1}$, $D$ approaches $2$ as $n\to\infty$. As discussed below, such solutions for large $n$ are distinct from those leading to the singular limit $D=2$ in the context of conformal field theory. 

\subsection{Conformally invariant equations}
The dilatation operator involved in scale invariance is part of the conformal group, which is infinite dimensional at $D=2$. It is of interest to examine the approach to $D=2$ from above in terms of our solutions where $D$ is noninteger. We first consider the formulation of the action in terms of conformal invariance \cite{DiFrancesco:1997nk} and show that there are two different families of solutions as we approach $D=2$ from above. 
\noindent{The conformally invariant action is}
\begin{equation}
\label{e29}
S=\int d^Dx\ \left(\tfrac{1}{2}\partial_\mu\phi\partial^\mu\phi-\lambda\phi^{2+
\frac{4}{D-2}}\right) 
\end{equation}
for $D\neq2$.  For $D=4$ we have a $\phi^4$ interaction, which is classically conformally invariant \cite{Erdmenger:1996yc, DiFrancesco:1997nk}. We are interested in solutions to \eqref{e4} as $D\to2$ from above. The classical Lagrangian is conformally invariant in dimension $D\neq2$. The scaling dimension of $\phi$ is $\tfrac{1}{2}(D-2)$
and $S$ in \eqref{e29} is invariant under the infinitesimal transformation $\phi\to\phi+\delta\phi$, where
\begin{equation}
\label{e30}
\delta\phi=v^\alpha\partial_\alpha\phi+\tfrac{D-2}{2D}\phi\partial_\alpha
v^\alpha;
\end{equation}
$v^\alpha$ is the conformal Killing vector in flat space 
\cite{Erdmenger:1996yc}.
We use our earlier methods for solving the radial form of the equation of motion for the action \eqref{e29}.

In our formalism \eqref{e29} represents a theory with $\epsilon=\frac{4}{D-2}$, so for the {\it conformally invariant} equation
\begin{equation}
\label{e31}
F(u)F'(u)-\tfrac{1}{4}(D-2)^2u+\tfrac{2\lambda D}{D-2} u^\frac{D+2}{D-2}=0.
\end{equation}
This fits into our framework of nonpoynomial solutions. However, we find additional solutions not covered by this case as $D$ approaches $2$. For $\epsilon=2n$ we have $D=2+\frac{2}{n}$ and, using the methods outlined above, we find that
\begin{equation}
\label{e32}
u(t)=[2\lambda n^2\sinh^2(t+\beta)]^{-
1/(2n)}\,,
\end{equation}
where $\beta$ is a constant of integration and $\lambda>0$.

We have thus found two new families of solutions as $D$ approachs $2$ from above. One is through conformally invariant solutions and the other is through solutions that can be real or complex depending on the sign of $\lambda$ and that are not conformally invariant. A discussion of existence theorems of solutions for the Euler-Lagrange equation for \eqref{e29} is given in~\cite{Mukhanov:2021rpp}.

\section{Conclusions} 
We have found many novel solutions to the saddle-point equations for scalar field theories. The possibilities for fractional interactions allow for new
deformations of Hermitian field theories that could be relevant for nonpolynomial field theory~\cite{Bender:1989vv} and also for \cPT-symmetric field theory \cite{R3b, Felski:2021evi, Croney:2023gwy, R3a, R4, PRD98.125003}. Furthermore, we have produced two different limiting procedures towards $D=2$. In principle, our solutions allow semiclassical evaluation of path integrals for unconventional field theories where canonical methods are cumbersome or impossible.

\section*{Acknowledgements}
SS thanks Wen-Yuan Ai for discussions on instantons. The work of SS and CMB is supported by the UK Engineering and Physical Sciences Research Council (grant EP/V002821/1). CMB is also supported by the Simons Foundation and the Alexander von Humboldt Foundation.

\appendix
\section*{Appendix}
In this Appendix we briefly describe two more examples of exact instanton solutions.
These examples are an additional illustration that the result of \eqref{e24} holds.\footnote{One should be aware of the 1-2-3-infinity fallacy in which one makes a general conclusion based on a few examples. A nice example of small-$n$ behaviour giving an incorrect answer for general $n$ is provided by the integral $\int_0^\infty dx\,f_n(x)$, where $f_n(x)=\prod_{k=1}^n\sin\left(\frac{x}{2k-1}\right)\Big/\frac{x}{2k-1}$. The integrals $\int_0^\infty dx\,f_n(x)=\frac{\pi}{2}$ for $n\le7$ but this relation fails at $n=8$. However, we have carefully checked that there are no such problems with the solutions presented in this paper.}
The case $\epsilon=6$ in \eqref{e9} leads to
\begin{equation}
\label{e18}
F(u)F'(u)+\left(d-\tfrac{5}{3}\right)F(u)-8\lambda u^7-\tfrac{1}{9}(3d-4)u=0.
\end{equation}
We substitute $F(u)=a_0+a_1u+a_2u^2+ a_3u^3+a_4u^4$ into \eqref{e18}, match powers of $u$, and get 8 equations in 5 unknowns:
\begin{align}
a_0a_1+\left(d-\tfrac{5}{3}\right)a_0&=0,\nonumber\\
a^2_1+2a_0a_2+\left(d-\tfrac{5}{3}\right)a_1+\tfrac{1}{9}(4-3d)&=0,\nonumber\\
3a_0a_3+3a_1a_2+\left(d-\tfrac{5}{3}\right)a_2&=0,\nonumber\\
4a_0a_4+\left(d-\tfrac{5}{3}\right)a_3+4a_1a_3+2a^2_2&=0,\nonumber\\
\left(d-\tfrac{5}{3}\right)a_4+5a_1a_4+5a_2a_3&=0,\nonumber\\
2a_2a_4+a^2_3&=0,\nonumber\\
a_3a_4&=0,\nonumber\\
a^2_4-2\lambda&=0.\label{e20h}
\end{align}
Solving \eqref{e20h} in {\it reverse} order, we obtain
$a_4=\pm\sqrt{2\lambda}$, $a_2=a_3=a_0=0$, $a_1=\left(\frac{1}{3}-\frac{d}{5}\right)$. From the second equation of \eqref{e20h} we deduce that either $d=0$ or $d=5/4$.

The case $\epsilon=8$ in \eqref{e9} leads to
\begin{equation}
\label{e21}
F(u)F'(u)+\left(d-\tfrac{3}{2}\right)F(u)-10\lambda u^9+\tfrac{1}{16}(5-4d)u=0.
\end{equation}
We substitute
$F(u)=a_0+a_1u+a_2u^2+a_3u^3+a_4u^4+a_5u^5$
into \eqref{e21} to obtain ten equations in six unknowns:
\begin{align*}
(a_1+d-\tfrac{3}{2})a_0&=0,\\
5-4d-24a_1+16a^2_1+32a_0a_2+16da_1&=0,
\\
a_1a_2+a_0a_3+\tfrac{1}{6}(2d-3)a_2&=0,\\
a^2_2+2a_1a_3+2a_0a_4+\tfrac{1}{4}(2d-3)a_3&=0,\\
a_2a_3+a_1a_4+a_0a_5+\tfrac{1}{10}(2d-3)a_4&=0,\\
a^2_3+2a_2a_4+2a_1a_5+\tfrac{1}{6}(2d-3)a_5&=0,\\
a_3a_4+a_2a_5&=0,\\
a^2_4+2a_3a_5&=0,\\
a_4a_5&=0,\\
a^2_5-2\lambda&=0.
\end{align*}
The solution to these equations follows the pattern of the solutions for $\epsilon=2,4,6$ above:
$$a_0=a_2=a_3=a_4=0,\quad
a_5=\pm\sqrt{2\lambda},\quad
a_1=\tfrac{1}{4}-\tfrac{d}{6}.$$
Hence, either $d=0$ or $d=6/5$.

\subsection*{Conflict of interest statement} There is no conflict of interest in regards this work.

\subsection*{Data availability statement} No new data were created or analysed in this study.

\bibliography{bibli.bib}
\end{document}